\documentclass[aps,prc,twocolumn,groupaddress,showpacs,nofootinbib,floatfix]{revtex4}

\usepackage{bm}
\usepackage{epsfig}
\usepackage{dcolumn}

\def\bea {\begin{eqnarray}}
\def\eea {\end{eqnarray}}
\def\be {\begin{equation}}
\def\ee {\end{equation}}
\def\ben{\begin{enumerate}}
\def\een{\end{enumerate}}
\def\bi{\begin{itemize}}
\def\ei{\end{itemize}}

\def\etal{{\it et al.}}

\def\F{{\cal F}}

\def\prl {Phys. Rev. Lett.\ }

\def\pl {Phys. Lett.\ }
\def\pr {Phys. Rev.\ }

\def\Vud{V_{\mbox{\scriptsize ud}}}

\def\hyphen{{\mbox{-}}}

\def\2p{|2p\rangle }
\def\4p2h{|4p\hyphen 2h\rangle }
\def\6p4h{|6p\hyphen 4h\rangle }

\def\overj2{\overline{\j}_2}

\begin{document} 

\title{Theoretical corrections and world data for the superallowed $ft$ values in the $\beta$
decays of $^{42}$Ti, $^{46}$Cr, $^{50}$Fe and $^{54}$Ni}

\author{I.S. Towner}
\email{towner@comp.tamu.edu}
\author{J.C. Hardy} 
\email{hardy@comp.tamu.edu}
\affiliation{Cyclotron Institute, Texas A\&M University, College Station, Texas 77843}

\date{\today}

\begin{abstract} 
Prompted by recent measurements, we have surveyed world data and calculated radiative and isospin-symmetry breaking corrections for the
superallowed $0^+ \rightarrow 0^+$ Fermi transitions from $^{42}$Ti, $^{46}$Cr, $^{50}$Fe and $^{54}$Ni.  This increases the number of such
transition with a complete set of calculated corrections from 20 to 23.  The results are compared with their equivalents for the mirror
superallowed transitions from $^{42}$Sc, $^{46}$V, $^{50}$Mn and $^{54}$Co.  The predicted $ft$-value asymmetries of these mirror pairs are
sensitive to the correction terms and provide motivation for improving measurement precision so as to be able to test the corrections.  To aid   
in that endeavor, we present a parameterization for calculating the $f$ values for the new transitions to $\pm$0.01\%.

\end{abstract} 


\maketitle

\section{Introduction}
\label{s:intro}

At regular intervals over more than four decades, we have published critical surveys of world data on superallowed
$0^+$$\rightarrow 0^+$ Fermi $\beta$ transitions and their impact on weak-interaction physics, with the last survey
appearing in February 2015 \cite{HT15}.  In all, 20 transitions were included in this most-recent survey, of which
18 had a complete set of data, comprising in each case the $Q_{EC}$ value, half-life and branching ratio. Of those 18,
all but 4 had been measured to high precision.  Our justification for including 20 cases, some of which were incomplete
or poorly known, was that we deemed these 20 cases to encompass all those that were likely to be accessible to precision
measurements in the near future.

By the time the survey was published, our prediction had already been proven wrong: In January 2015, Molina \etal\,\cite{Mo15}
reported a measurement of the half-lives and Gamow-Teller branching ratios for the $\beta$ decays of $^{42}$Ti, $^{46}$Cr,
$^{50}$Fe and $^{54}$Ni.  Although the $^{42}$Ti transition was included in our survey, those of $^{46}$Cr, $^{50}$Fe and
$^{54}$Ni were not.  In fact, the $Q_{EC}$ values for the three latter transitions are still poorly known and even the new
measurements of the half-lives and branching ratios have yet to reach the precision required to contribute meaningfully to
any standard-model tests.  Nevertheless, Molina \etal\, have convincingly demonstrated that these nuclei are indeed
accessible and potentially amenable to more precise measurements.  

This report is intended as an addendum to our 2015 survey \cite{HT15}, in which we extend the same evaluation of world data
to the three new superallowed transitions and, more importantly, evaluate all the correction terms that are required to 
understand the results.  At the same time, we take the opportunity to update results for $^{42}$Ti to incorporate the new
information.

A $\beta$ transition is characterized by its $ft$ value, where $f$ is the statistical rate function and $t$ is its partial
half-life.  Three experimental quantities are required to establish the $ft$ value: The total decay energy, $Q_{EC}$, is
required to calculate $f$; and the half-life, $t_{1/2}$, and the branching ratio, $R$, combine as follows to produce the
partial half-life:  
\be 
t = \frac{t_{1/2}}{R} \left ( 1 + P_{EC} \right ) .
\label{partial}
\ee
Here $P_{EC}$ is a small correction to account for competition from electron capture.

To the $ft$ value, two theoretical corrections are applied to produce a corrected $\F t$ value, defined as
\bea
\F t & = & ft (1 + \delta_R ) ( 1 - \delta_C)
\nonumber \\
& = & ft (1 + \delta_R^{\prime})(1 - \delta_C + \delta_{NS}) .
\label{Ft}
\eea
Here $\delta_R$ is the nucleus-dependent part of the radiative correction, also called the ``outer" radiative correction,
and $\delta_C$ is an isospin-symmetry-breaking correction.  It is convenient to subdivide $\delta_R$ further as $\delta_R
= \delta_R^{\prime} + \delta_{NS}$ and, since these quantities are small, rearrange the equation to the form displayed on the
second line of Eq.~(\ref{Ft}), which is correct to first order in these corrections.  This rearrangement places the
nuclear-structure-dependent corrections together in the combination $\delta_C - \delta_{NS}$.

In what follows, we begin with a survey of world data for the superallowed $\beta$-decay branches of the $T_z=-1$ nuclei, $^{42}$Ti,
$^{46}$Cr, $^{50}$Fe and $^{54}$Ni, from which the $ft$ values are obtained.  Next we calculate the correction terms
$\delta_R^{\prime}$, $\delta_{NS}$ and $\delta_C$, and hence obtain $\F t$ values for these 4 cases.  These are then compared
with results for the well-known superallowed decay branches from the $T_z=0$ nuclei $^{42}$Sc, $^{46}$V, $^{50}$Mn and $^{54}$Ni,
which are their mirror transitions.  Finally we use the calculated correction terms to predict the ratio of $ft$ values for each
of the four pairs of mirror transitions.  When the precision of world data is improved for the $T_z=-1$ cases, this will provide a
stringent test of the correction terms \cite{TH10}.

Our focus here is on providing information that will be useful to experimenters when such improvements have been achieved.  In
that context, we also tabulate the parameters needed to calculate easily the $f$ values for the three new transitions -- from $^{46}$Cr,
$^{50}$Fe and $^{54}$Ni -- to high precision ($\pm$0.01\%), which will be important once more precise $Q_{EC}$ values are known.  These
parameters supplement those given in Ref.\,\cite{TH15} for the 20 previously surveyed transitions.

\begin{table}[t]
\caption {Measured results from which the decay transition energies, $Q_{EC}$, have been derived for the superallowed
$\beta$-decays of four $T_z=-1$ nuclei.  In all cases only a single useful measurement has been made of each quantity.  The lines giving the superallowed
$Q_{EC}$ values themselves are in bold print. Where no reference is given, the $Q_{EC}$ value was determined from the
difference between the measured parent and daughter mass excesses.  (See Table~\ref{ref} for the correlation between the
alphanumeric reference code used in this table and the actual reference numbers.)}
\label{QEC}
\begin{ruledtabular}
\begin{tabular}{llll} \\ [-3mm]

\multicolumn{2}{c}{Parent/Daughter}
 & Property\footnotemark[1] 
 & Measured Energies used \\

   \multicolumn{2}{c}{nuclei} &
 & to determine $Q_{EC}$ (keV) \\[1mm]

\hline &&& \\[-2mm]

~~ $^{42}$Ti & $^{42}$Sc & $\bm{Q_{EC}(sa)}$ & ~{\bf 7016.83 $\pm$ 0.25} ~[Ku09]  \\ [2mm]
~~ $^{46}$Cr & $^{46}$V & ME($p$) & ~~~~-29474 $\pm$ 20 ~[Zi72]  \\ 
 & & ME($d$) & $\,$-37074.55 $\pm$ 0.32\footnotemark[2] \\
 & & $\bm{Q_{EC}(sa)}$ & ~~~~~~{\bf 7600 $\pm$ 20} \\ [2mm]
~~ $^{50}$Fe & $^{50}$Mn & ME($p$) & ~~~~$\,$-34489 $\pm$ 60 ~[Tr77]  \\ 
 & & ME($d$) & $\,$-42627.25 $\pm$ 0.90\footnotemark[2] \\
 & & $\bm{Q_{EC}(sa)}$ & ~~~~~~{\bf 8139 $\pm$ 60} \\ [2mm]
~~ $^{54}$Ni & $^{54}$Co & ME($p$) & ~~~~$\,$-39223 $\pm$ 50 ~[Tr77]  \\ 
 & & ME($d$) & $\,$-48009.52 $\pm$ 0.56\footnotemark[2] \\
 & & $\bm{Q_{EC}(sa)}$ & ~~~~~~{\bf 8787 $\pm$ 50} \\ [-2mm]

\footnotetext[1]{Abbreviations used in this column are as follows: 
``$sa$", superallowed transition; ``$p$", parent; ``$d$", daughter; and ``$ME$", mass excess;  Thus, for
example, ``$Q_{EC}(sa)$" signifies the $Q_{EC}$-value for the superallowed transition, and ``$ME(d)$", the mass
excess of the daughter nucleus.}
\footnotetext[2]{Result obtained from the $Q_{EC}$ value for the superallowed decay of the daughter $d$, which
appears in Ref.$\,$\cite{HT15}, combined with the mass of the grand-daughter taken from [Wa12].}
\end{tabular}
\end{ruledtabular}
\end{table}

\begin{table*}
\caption{Half-lives, $t_{1/2}$, of four $T_z=-1$ superallowed $\beta$-emitters.  (See Table~\ref{ref} for the correlation
between the   alphabetical reference code used in this table and the actual reference numbers.)
\label{t1/2}}
\begin{ruledtabular}
\begin{tabular}{llllllll}
Parent   
 & \multicolumn{4}{c}{Measured half-lives, $t_{1/2}$ (ms)}
 & \multicolumn{1}{c}{}
 & \multicolumn{2}{c}{Average value} \\[1mm]
\cline{2-5} 
\cline{7-8} \\[-2mm]
nucleus 
 & \multicolumn{1}{c}{1}
 & \multicolumn{1}{c}{2} 
 & \multicolumn{1}{c}{3} 
 & \multicolumn{1}{c}{4} 
 & \multicolumn{1}{c}{}
 & \multicolumn{1}{c}{$t_{1/2}$ (ms)} 
 & \multicolumn{1}{c}{scale} \\[1mm]
\hline  &&&&&&& \\[-2mm]

~~ $^{42}$Ti & ~~\,202 $\pm$ 5 ~~[Ga69] & 208.14 $\pm$ 0.45 [Ku09] & 211.7 $\pm$ 1.9 [Mo15] & 209.5 $\pm$ 5.2 [Mo15] & & ~~208.29 $\pm$ 0.79 & 1.8 \\
~~ $^{46}$Cr & 224.3 $\pm$ 1.3 [Mo15] & ~\,223.9 $\pm$ 9.9 ~\,[Mo15] & & & & ~~~\,224.3 $\pm$ 1.3 & 1.0 \\
~~ $^{50}$Fe & 152.1 $\pm$ 0.6 [Mo15] & ~\,150.1 $\pm$ 2.9 ~\,[Mo15] & & & & ~~152.02 $\pm$ 0.59 & 1.0 \\
~~ $^{54}$Ni & 114.2 $\pm$ 0.3 [Mo15] & ~\,114.3 $\pm$ 1.8 ~\,[Mo15] & & & & ~~114.20 $\pm$ 0.30 & 1.0 \\[1mm]
\end{tabular}
\end{ruledtabular}
\end{table*}

\section{Experimental Data}
\label{s:expt}

\begin{table*}
\caption{Measured results from which the branching ratios, R, have been derived for superallowed $\beta$-transitions
from four $T_z=-1$ nuclei.  The lines giving the average superallowed branching ratios themselves are in bold print.
( See Table~\ref{ref} for the correlation between the alphabetical reference code used in this table and the actual
reference numbers.)
\label{R}}
\begin{ruledtabular}
\begin{tabular}{llllllll}
\multicolumn{2}{c}{Parent/Daughter}
 & Daughter state
 & \multicolumn{2}{c}{Measured Branching Ratio, R (\%)}
 & \multicolumn{1}{c}{}
 & \multicolumn{2}{c}{Average value} \\[1mm]
\cline{4-6} 
\cline{7-8} \\[-1mm]
   \multicolumn{2}{c}{nuclei}
 & \multicolumn{1}{c}{$E_x$ (MeV)}
 & \multicolumn{1}{c}{1}
 & \multicolumn{1}{c}{2} 
 & \multicolumn{1}{c}{} 
 & \multicolumn{1}{c}{R (\%)} 
 & \multicolumn{1}{c}{scale} \\[1mm]

\hline
& & & & & & & \\[-2mm]

~~ $^{42}$Ti & $^{42}$Sc & 0.611 & 51.1 $\pm$ 1.1 [Ku09] & 55.9 $\pm$ 3.6 [Mo15] & & ~51.5 $\pm$ 1.3 & 1.3 \\
 & & {\bf gs} & & & & {\bf 48.1 $\pm$ 1.4\footnotemark[1]} & \\[2mm]
~~ $^{46}$Cr & $^{46}$V & 0.994 & 21.6 $\pm$ 5.0 [On05] & 13.9 $\pm$ 1.0 [Mo15] & & ~14.2 $\pm$ 1.4 & 1.5 \\
 & & {\bf gs} & & & & {\bf 76.7 $\pm$ 2.3\footnotemark[1]} & \\[2mm]
~~ $^{50}$Fe & $^{50}$Mn & 0.651 & 22.5 $\pm$ 1.4 [Mo15] & & & ~22.5 $\pm$ 1.4 &  \\
 & & {\bf gs} & & & & {\bf 74.3 $\pm$ 1.6\footnotemark[1]} & \\[2mm]
~~ $^{54}$Ni & $^{54}$Co & 0.937 & 22.4 $\pm$ 4.4 [Re99] & 19.8 $\pm$ 1.2 [Mo15] & & ~19.9 $\pm$ 1.2 & 1.0 \\
 & & {\bf gs} & & & & {\bf 78.9 $\pm$ 1.2\footnotemark[1]} & \\[-3mm]

\footnotetext[1]{Result also incorporates data from Table~\ref{BDG}}.

\end{tabular}
\end{ruledtabular}
\end{table*}

\begin{table}
\caption{Relative intensities of $\beta$-delayed $\gamma$-rays in the superallowed $\beta$-decay daughters.  These data
are used to determine the branching ratios presented in Table~\ref{R}.  (See Table~\ref{ref}
for the correlation between the alphabetical reference code used in this table and the actual reference numbers.)
\label{BDG}}
\begin{ruledtabular}
\begin{tabular}{lcll}
\multicolumn{2}{c}{Parent/Daughter}
 & \multicolumn{1}{c}{daughter}
 & \multicolumn{1}{c}{Measured} \\

   \multicolumn{2}{c}{nuclei}
 & \multicolumn{1}{c}{ratios\footnotemark[1]}
 & \multicolumn{1}{c}{$\gamma$-ray Ratio}\\[1mm]

\hline
& & & \\[-2mm]

$^{42}$Ti & $^{42}$Sc & $\gamma_{total}/\gamma_{611}$ & ~\,0.0073 $\pm$ 0.0011\footnotemark[2] \,[Mo15] \\
$^{46}$Cr & $^{46}$V & $\gamma_{total}/\gamma_{994}$ & ~~~0.642 $\pm$ 0.026 ~~~[Mo15] \\
$^{50}$Fe & $^{50}$Mn & $\gamma_{total}/\gamma_{651}$ & ~~~0.158 $\pm$ 0.015 ~~~[Mo15] \\
$^{54}$Ni & $^{54}$Co & $\gamma_{total}/\gamma_{937}$ & ~\,0.0576 $\pm$ 0.0043 ~\,[Mo15] \\ [-2mm]

\footnotetext[1]{$\gamma$-ray intensities are denoted by $\gamma_{E}$, where $E$ is the $\gamma$-ray energy in keV. The
notation $\gamma_{total}$ appearing in a numerator denotes the sum of all $\beta$-delayed $\gamma$ rays feeding the
daughter ground state, excluding the strongest $\gamma$ ray, which is identified in the denominator.}
\footnotetext[2]{This result replaces the result appearing in our 2015 survey \cite{HT15}, which came from [Ga69] and [En90]. The
2223-keV $\gamma$ ray identified in [Ga69] as originating from $^{42}$Ti decay evidently originated from a contaminant since
it was not observed in [Mo15].}
\end{tabular}
\end{ruledtabular}
\end{table}

\begin{table}
\caption{Reference key, relating alphabetical reference codes used in Tables~\ref{QEC}-\ref{BDG} to the actual reference numbers.
\label{ref}}
\vskip 1mm
\begin{ruledtabular}
\begin{tabular}{lclclcl}
  Table & Reference & Table & Reference & Table & Reference \\ 
  code & number & code & number & code & number \\
\hline \\[-2mm]
  En90  & \cite{En90}  &
  Ga69  & \cite{Ga69}  &
  Ku09  & \cite{Ku09}  \\
  Mo15  & \cite{Mo15}  &
  On05  & \cite{On05}  &
  Re99  & \cite{Re99}  \\
  Tr77  & \cite{Tr77}  &
  Wa12  & \cite{Wa12}  &
  Zi72  & \cite{Zi72}  \\
      
\end{tabular}
\end{ruledtabular}
\end{table}

We surveyed world data using exactly the same methods as in our 2015 survey \cite{HT15} and, for consistency, we present the
results here in a similar tabular format, even though relatively few references are involved. The $Q_{EC}$ values appear in
Table~\ref{QEC}, the half-lives in Table~\ref{t1/2} and the branching ratios in Table~\ref{R}.  Since the branching ratios
for the decay of a $T_z=-1$ nucleus depends on a complete analysis of its spectrum of $\beta$-delayed $\gamma$ rays, we give
in Table~\ref{BDG} the relative intensities of the $\gamma$ rays for all four cases.  As in the survey, each datum appearing
in the tables is attributed to its original journal reference $via$ an alphanumeric code made up of the initial two letters
of the first author's name and the two last digits of the publication date.  These codes are correlated with the actual
reference numbers, Refs.\,\cite{En90}-\cite{Zi72}, in Table~\ref{ref}.

Several remarks can be made concerning the contents of the tables.  Table~\ref{QEC} shows that the $Q_{EC}$ value for $^{42}$Ti
decay has been directly measured quite recently; as reported in Ku09 \cite{Ku09}, this was done with a Penning trap and is rather
precisely known.  The other three $Q_{EC}$ values in the table had to be derived as differences between separately measured
parent and daughter masses.  Furthermore, all three parent masses were measured about 40 years ago, either from a reaction
excitation function (see Zi72~\cite{Zi72}) or from the $Q$-values of ($^4$He, $^8$He) reactions (see Tr77~\cite{Tr77}) and have
large uncertainties by today's standards.  Note also that the result for $^{42}$Ti is the same as appeared in our 2015 survey
\cite{HT15}.

In Tables~\ref{t1/2}, \ref{R} and \ref{BDG} the survey results for $^{42}$Ti have been updated for new data from Mo15 \cite{Mo15}.  In
particular, the branching-ratio result has been changed significantly since Mo15 did not observe a $\beta$-delayed $\gamma$ ray
that had been attributed to $^{42}$Ti.  This is explained fully in footnote b of Table~\ref{BDG}.

\begin{table*}
\caption{Results derived from Tables \ref{QEC}-\ref{BDG} for the four superallowed Fermi beta decays from $T_z=-1$ nuclei.  Also shown for
comparison are the equivalent results for their mirror transitions from $T_z=0$ nuclei; these are taken from Ref.~\cite{HT15}.
\label{t:tabb}}
\begin{ruledtabular}
\begin{tabular}{rrrrrrrr}
& & & & & & & \\[-3mm]
\multicolumn{1}{r}{Parent} & &
\multicolumn{1}{r}{$P_{EC}$} &
\multicolumn{1}{r}{Partial half life} & & & & \\
\multicolumn{1}{r}{nucleus} & 
\multicolumn{1}{c}{$f$} & 
\multicolumn{1}{r}{$( \% )$} & 
\multicolumn{1}{c}{$t$(ms)} & 
\multicolumn{1}{c}{$ft$(s)} & 
\multicolumn{1}{r}{$\delta_R^{\prime}( \% )$} & 
\multicolumn{1}{r}{$\delta_C - \delta_{NS} ( \% )$} & 
\multicolumn{1}{c}{~~$\F t$ (s)} \\[1mm]
\hline
& & & & & & & \\[-2mm]
\multicolumn{1}{l}{$T_z=-1$} &&&&&&&\\
$^{42}$Ti & $ 7130.5 \pm 1.4~\; $ & 0.087 & $433 \pm 12~~\,$ & $3090 \pm 88~\,$ &
1.427 & $1.195 \pm 0.066 $ & $ 3096 \pm 88~\, $ \\
$^{46}$Cr & $10660 \pm 150~ $ & 0.092 & $ 292.6 \pm 9.1~\, $ & $3120 \pm 110 $ &
1.420 & $ 0.935 \pm 0.090 $ & $3130 \pm 110 $ \\
$^{50}$Fe & $14950 \pm 600~ $ & 0.100 & $ 204.8 \pm 4.5~\, $ & $ 3060 \pm 140 $ &
1.439 & $ 0.815 \pm 0.053 $ & $3080 \pm 140 $ \\
$^{54}$Ni & $ 21850 \pm 670~ $ & 0.104 & $144.9 \pm 2.3~\, $ & $ 3170 \pm 110 $ &
1.430 & $ 0.955 \pm 0.070 $ & $ 3180 \pm 110 $ \\[1mm]
\multicolumn{1}{l}{$T_z=0$} &&&&&&&\\
$^{42}$Sc & $ 4472.23 \pm 1.15 $ & 0.099 & $ 681.44 \pm 0.26 $ & $ 3047.5 \pm 1.4\, $ &
1.453 & $ 0.655 \pm 0.050 $ & $ 3071.6 \pm 2.1 $ \\
$^{46}$V  & $ 7209.25 \pm 0.54 $ & 0.101 & $ 423.113^{+0.053}_{-0.056}\, $ & $ 3050.32 ^{+0.44}_{-0.46}\, $ &
1.445 & $ 0.655 \pm 0.063 $ & $ 3074.1 \pm 2.0 $ \\
$^{50}$Mn & $ 10745.97 \pm 0.50 $ & 0.107 & $ 283.68 \pm 0.11 $ & $ 3048.4 \pm 1.2 $ &
1.444 & $ 0.705 \pm 0.034 $ & $ 3070.6 \pm 1.6 $ \\
$^{54}$Co & $15766.7 \pm 2.9~\, $ & 0.111 & $ 193.493 ^{+0.063}_{-0.086}\, $ & $ 3050.7 ^{+1.1}_{-1.5}~\, $ &
1.443 & $ 0.805 \pm 0.068 $ & $ 3069.8 ^{+2.4}_{-2.6}~\, $ \\[1mm]
\end{tabular}
\end{ruledtabular}
\end{table*}

With the input data now settled, we can derive the $ft$ values for the four superallowed transitions from the $T_z=-1$ nuclei,
$^{42}$Ti, $^{46}$Cr, $^{50}$Fe and $^{54}$Ni.  The results appear in the top four rows of Table~\ref{t:tabb}, where we give
the statistical rate functions, $f$, the electron-capture fractions, $P_{EC}$, the partial half-lives, $t$, obtained with
Eq.~(\ref{partial}), and finally the $ft$ values.  To facilitate later mirror comparisons, we also give the same information
for the four mirror transitions from the $T_z=0$ nuclei, $^{42}$Sc, $^{46}$V, $^{50}$Mn and $^{54}$Co.  These are identical
to the results that appear in Table IX of Ref.~\cite{HT15}. 

The next step is to determine the theoretical correction terms $\delta_R^{\prime}$, $\delta_{NS}$ and $\delta_C$.  Their derivation
is described in the next section.

\section{Theoretical corrections}
\label{s:theo}

\subsection{Outer radiative correction, $\delta_R$}
\label{ss:outer}

As noted already, the nucleus-dependent outer radiative correction $\delta_R$ is conveniently divided into two components,
\be
\delta_R = \delta_R^{\prime} + \delta_{NS} .
\label{dr}
\ee
The first comprises the bremsstrahlung and low-energy part of the $\gamma W$-box graphs and is a standard QED calculation that depends
only on the electron's energy and the charge, $Z$, of the daughter nucleus.

The calculation of $\delta_R^{\prime}$ can be further broken down into four contributions \cite{TH08}:
\be
\delta_R^{\prime} = \frac{\alpha}{2 \pi} \left [
\overline{g}(E_m) + \delta_2 + \delta_3 + \delta_{\alpha^2}
\right ] .
\label{dRprime}
\ee
The leading order-$\alpha$ term contains the function $\overline{g}(E_m)$: It is the average over the $\beta$ energy spectrum of the
function $g(E,E_m)$, originally defined by Sirlin \cite{Si67}.  Here $E$ is the total electron energy in the $\beta$-decay transition
and $E_m$ is its maximum value. The next two terms in Eq.~(\ref{dRprime}), $\delta_2$ and $\delta_3$, represent corrections to order
$Z \alpha^2$ and $Z^2 \alpha^3$ respectively.  The last term is a recently-added contribution \cite{TH08} that gives a correction
to order $\alpha^2$.

Results for all four terms and their sums are recorded in Table~\ref{t:tab1} for the superallowed decays of $^{42}$Ti, $^{46}$Cr,
$^{50}$Fe and $^{54}$Ni, as well as for their mirror superallowed transitions.  The differences in the radiative corrections for each
pair of mirror transitions are given in the last four lines of the table. They are very small.

\begin{table*}
\caption{Calculated transition-dependent radiative corrections $\delta_R^{\prime}$ in percent units, and their component contributions.
As explained in the text, no uncertainty is given.  The results for $^{46}$Cr, $^{50}$Fe and $^{54}$Ni are presented here for the first
time; the results for the other cases are the same as those appearing in Ref.~\cite{TH08}.  The last four lines give the differences in
radiative-correction terms for the designated mirror transitions.
\label{t:tab1}}
\begin{ruledtabular}
\begin{tabular}{rrrrrr}
& & & & & \\[-3mm]
\multicolumn{1}{r}{Parent} &
\multicolumn{1}{r}{$\frac{\alpha}{2 \pi} \overline{g}(E_m)$} &
\multicolumn{1}{r}{$\frac{\alpha}{2 \pi} \delta_2$ } &
\multicolumn{1}{r}{$\frac{\alpha}{2 \pi} \delta_3$ } &
\multicolumn{1}{r}{$\frac{\alpha}{2 \pi} \delta_{\alpha^2}$ } &
\multicolumn{1}{r}{$\delta_R^{\prime} $ } \\
\multicolumn{1}{r}{nucleus} & & & & & \\[1mm]
\hline
& & & & & \\[-2mm]
\multicolumn{1}{l}{$T_z=-1$} &&&&&\\
$^{42}$Ti & 0.9051 & 0.4556 & 0.0501 & 0.0160 & 1.4269 \\
$^{46}$Cr & 0.8745 & 0.4734 & 0.0567 & 0.0154 & 1.4200 \\
$^{50}$Fe & 0.8489 & 0.5077 & 0.0675 & 0.0148 & 1.4390 \\
$^{54}$Ni & 0.8203 & 0.5205 & 0.0747 & 0.0144 & 1.4299 \\[1mm]
\multicolumn{1}{l}{$T_z=0$} &&&&&\\
$^{42}$Sc & 0.9392 & 0.4507 & 0.0467 & 0.0166 & 1.4533 \\
$^{46}$V  & 0.9031 & 0.4720 & 0.0539 & 0.0159 & 1.4448 \\
$^{50}$Mn & 0.8728 & 0.4942 & 0.0620 & 0.0153 & 1.4444 \\
$^{54}$Co & 0.8440 & 0.5134 & 0.0707 & 0.0147 & 1.4427 \\[3mm]
$^{42}$Sc $-$ $^{42}$Ti & $ 0.0341$ & $-0.0049$ & $-0.0034$ & $ 0.0006$ & $ 0.0264$ \\
$^{46}$V  $-$ $^{46}$Cr & $ 0.0286$ & $-0.0014$ & $-0.0028$ & $ 0.0005$ & $ 0.0248$ \\
$^{50}$Mn $-$ $^{50}$Fe & $ 0.0239$ & $-0.0135$ & $-0.0055$ & $ 0.0005$ & $ 0.0054$ \\
$^{54}$Co $-$ $^{54}$Ni & $ 0.0237$ & $-0.0071$ & $-0.0040$ & $ 0.0003$ & $ 0.0128$ \\[1mm]
\end{tabular}
\end{ruledtabular}
\end{table*}

No uncertainties on $\delta_R^{\prime}$ are listed in Table~\ref{t:tab1}.  This issue has been discussed in our recent survey
\cite{HT15}, where it is argued that the uncertainty on $\delta_R^{\prime}$ should be treated as a systematic, rather than a
statistical one.  We take the magnitude of the uncertainty to be one-third the contribution of the $Z^2 \alpha^3$ term but
apply it only to the final average $\overline{\F t}$ value, so that its influence is not reduced by statistical averaging.

The second component of the outer radiative correction, $\delta_{NS}$, recognizes that the $\gamma W$-box graph includes
situations in which the $\gamma$-nucleon interaction in the nucleus does not involve the same nucleon as the one participating
in the $W$-nucleon interaction.  When this happens, two distinct nucleons are actively involved and a detailed shell-model
calculation is required to evaluate $\delta_{NS}$.  Being nuclear-structure dependent, there is some uncertainty in the
result, but fortunately $\delta_{NS}$ is smaller in magnitude than $\delta_R^{\prime}$ so this is not a serious impediment.
Our strategy has always been to mount several shell-model calculations with different effective interactions from the
literature, adopt an average value of $\delta_{NS}$ from the results for each transition, and assign an uncertainty that
embraces the range of results obtained.  We follow that approach here too.  We also use exactly the same sets of effective
interactions that we used in Ref.~\cite{TH08}, where they are described in more detail and fully referenced.

\begin{table*}
\caption{Calculated nuclear-structure-dependent radiative correction $\delta_{NS}$.  The four components that are summed to give
$C_{NS}^{\rm quenched}$ characterize the four electromagnetic couplings:  os = orbital isoscalar, ss = spin isoscalar, ov = orbital
isovector, and sv = spin isovector.  The table gives one sample shell-model result, while the adopted value gives an average over
several different shell-model calculations, with an uncertainty that embraces the range. The last four lines give the difference
in radiative corrections for mirror transitions.  Note that the uncertainties of the mirror differences in $\delta_{NS}$ were
not determined from the uncertainties on the two contributing $\delta_{NS}$ values but were independently evaluated to cover the spread
in the calculated {\it differences}.  
\label{t:tab2}}
\begin{ruledtabular}
\begin{tabular}{rrrrrrrrr}
& & & & & & & & \\[-3mm]
\multicolumn{1}{r}{Parent} &
\multicolumn{5}{c}{$C_{NS}^{\rm quenched}$ } &
\multicolumn{1}{r}{$(q-1) C_{\rm Born}^{\rm free}$ } &
\multicolumn{1}{r}{$\delta_{NS}(\%)$ } &
\multicolumn{1}{r}{$\delta_{NS}(\%)$ } \\[1mm]
\cline{2-6}
\\[-2mm]
\multicolumn{1}{r}{nucleus} & 
\multicolumn{1}{r}{os} &
\multicolumn{1}{r}{ss} &
\multicolumn{1}{r}{ov} &
\multicolumn{1}{r}{sv} &
\multicolumn{1}{r}{total} & & & 
\multicolumn{1}{r}{adopted} \\[1mm]
\hline
& & & & \\[-2mm]
\multicolumn{1}{l}{$T_z=-1$} &&&&&&&&\\
$^{42}$Ti & $-0.019$ & $-0.160$ & $-0.207$ & $-0.388$
 & $-0.774$ & $-0.241$ & $-0.236$ & $-0.235(20)$ \\
$^{46}$Cr & $-0.004$ & $-0.197$ & $-0.099$ & $-0.198$ 
& $-0.498$ & $-0.248$ & $-0.173$ & $-0.175(20)$ \\
$^{50}$Fe & $-0.009$ & $-0.185$ & $-0.104$ & $-0.153$ 
& $-0.451$ & $-0.254$ & $-0.164$ & $-0.155(20)$ \\
$^{54}$Ni & $-0.012$ & $-0.180$ & $-0.133$ & $-0.203$ 
& $-0.528$ & $-0.261$ & $-0.183$ & $-0.165(20)$ \\[1mm]
\multicolumn{1}{l}{$T_z=0$} &&&&&&&&\\
$^{42}$Sc & $-0.019$ & $-0.160$ & 0.207 & 0.388 
& 0.416 & $-0.241$ & 0.041 & 0.035(20) \\
$^{46}$V  & $-0.004$ & $-0.197$ & 0.099 & 0.198 
& 0.096 & $-0.248$ & $-0.035$ & $-0.035(10)$ \\
$^{50}$Mn & $-0.009$ & $-0.185$ & 0.104 & 0.153 
& 0.063 & $-0.254$ & $-0.044$ & $-0.040(10)$ \\
$^{54}$Co & $-0.012$ & $-0.180$ & 0.133 & 0.203 
& 0.144 & $-0.261$ & $-0.027$ & $-0.035(10)$ \\[3mm]

$^{42}$Sc $-$ $^{42}$Ti & 0.000 & 0.000 & $ 0.414$ & $ 0.776$ 
& $ 1.190$ & 0.000 & $ 0.276$ & $ 0.270(30)$ \\
$^{46}$V  $-$ $^{46}$Cr & 0.000 & 0.000 & $ 0.198$ & $ 0.396$ 
& $ 0.594$ & 0.000 & $ 0.138$ & $ 0.140(10)$ \\
$^{50}$Mn $-$ $^{50}$Fe & 0.000 & 0.000 & $ 0.208$ & $ 0.306$ 
& $ 0.514$ & 0.000 & $ 0.119$ & $ 0.115(20)$ \\
$^{54}$Co $-$ $^{54}$Ni & 0.000 & 0.000 & $ 0.266$ & $ 0.406$ 
& $ 0.672$ & 0.000 & $ 0.156$ & $ 0.130(30)$ \\[1mm]
\end{tabular}
\end{ruledtabular}
\end{table*}

The calculation of $\delta_{NS}$ is based on the formula
\be
\delta_{NS} = \frac{\alpha}{\pi} \left [ C_{NS}^{\rm quenched} +
(q-1) C_{\rm Born}^{\rm free} \right ] ,
\label{dNS}
\ee
where the component terms are defined and discussed in Ref.~\cite{TH02}.  We use quenched electroweak vertices in the nucleus
\cite{To94}, so $q$ represents the quenching factor by which the product of the weak and electromagnetic coupling constants is
reduced in the medium relative to its free-nucleon value.  Detailed results are given in columns 2-5 of Table~\ref{t:tab2},
where we show contributions to $C_{NS}^{\rm quenched}$ from the various components of the electromagnetic interaction: orbital
isoscalar (os), spin isoscalar (ss), orbital isovector (ov), and spin isovector (sv).  Note that the spin contributions are
larger than the orbital contributions.

An even more interesting observation from Table~\ref{t:tab2} is that the isoscalar and isovector contributions to $\delta_{NS}$
are in phase when the decaying nucleus has $T_z = -1$ and out of phase when it has $T_z = 0$.  This leads to larger corrections
for transitions from the $T_z = -1$ nuclei than for those from the $T_z =0$ ones.  As is made clear by the differences in
mirror $\delta_{NS}$ values shown in the bottom four lines of the last column in Table~\ref{t:tab2}, this effect creates an
asymmetry of between 0.1 and 0.3\%.  This asymmetry would of course contribute to the expected mirror asymmetry in the
experimental $ft$ values and, since current experiments aim at 0.1\% precision, this effect is just  at the edge of detectability.

\subsection{Isospin-symmetry-breaking correction, $\delta_C$}
\label{ss:isb}

The isospin-symmetry-breaking correction is defined as the reduction in the square of the Fermi matrix element, $|M_F|^2$, from
its symmetry-limit value, $|M_F^0|^2$.  Thus,
\be
|M_F|^2 = |M_F^0|^2 (1 - \delta_C) .
\label{dC}
\ee
For calculational convenience, we separate $\delta_C$ into two components \cite{TH08,HT15}
\be
\delta_C = \delta_{C1} + \delta_{C2} .
\label{c1c2}
\ee
The idea is that $\delta_{C1}$ follows from a tractible shell-model calculation that does not include significant nodal mixing,
while $\delta_{C2}$ corrects for the nodal mixing that would be present if the shell-model space were much larger.

For $\delta_{C1}$, a modest shell-model space (usually one major oscillator shell) is employed, in which Coulomb and other
charge-dependent terms have been added to the charge-independent effective Hamiltonian customarily used for the shell model.
However, the most-important Coulomb force is long range and its influence in configuration space extends much further than a
single major oscillator shell.  The principal impact of multi-shell mixing is to change the radial wave function of the proton
through mixing with radial functions that have more nodes.  In the $\beta$-decay matrix element, $M_F$, there is an overlap
between the radial functions of the proton and neutron that participate in the transition, and it is the reduction from unity
of the overlap integral that leads to the correction $\delta_{C2}$.

\begin{table*}
\caption{Shell-model calculation of the isospin-symmetry-breaking correction, $\delta_{C1}$. The table gives one sample
shell-model result, while the adopted value gives an average over several different shell-model calculations, with an
uncertainty that embraces the range.  The results for $^{46}$Cr, $^{50}$Fe and $^{54}$Ni are presented here for the first
time; the results for the other cases are the same as those appearing in Ref.~\cite{TH08}. The last four lines give the
difference in isospin-symmetry-breaking corrections for mirror transitions.  Note that the uncertainties of the mirror
differences in $\delta_{C1}$ were not determined from the uncertainties on the two contributing $\delta_{C1}$ values but were
independently evaluated to cover the spread in the calculated {\it differences}.  
\label{t:tab3}}
\begin{ruledtabular}
\begin{tabular}{rddddd}
& & & & & \\[-3mm]
\multicolumn{1}{r}{Parent} &
\multicolumn{1}{r}{$E_x(0^+)$ } &
\multicolumn{1}{r}{$E_x(0^+)$ } &
\multicolumn{1}{r}{$\delta_{C1}(\%)$ } &
\multicolumn{1}{r}{$\delta_{C1}(\%)$ } &
\multicolumn{1}{r}{$\delta_{C1}(\%)$ } \\
\multicolumn{1}{r}{nucleus} &  
\multicolumn{1}{r}{expt} &  
\multicolumn{1}{r}{SM} &  
\multicolumn{1}{r}{unscaled} &  
\multicolumn{1}{r}{scaled} &  
\multicolumn{1}{r}{adopted} \\[1mm]
\hline
& & & & & \\[-2mm]
\multicolumn{1}{l}{$T_z=-1$} &&&&&\\
$^{42}$Ti & 1.84 & 3.16 & 0.038 & 0.113 & 0.105(20) \\
$^{46}$Cr & 3.57\footnotemark[1] & 4.86 & 0.012 & 0.023 & 0.045(20) \\
$^{50}$Fe & 3.69 & 3.62 & 0.021 & 0.020 & 0.025(20) \\
$^{54}$Ni & 2.56 & 2.26 & 0.030 & 0.023 & 0.065(30) \\[1mm]
\multicolumn{1}{l}{$T_z=0$} &&&&&\\
$^{42}$Sc & 3.30\footnotemark[1] & 5.05 & 0.007 & 0.017 & 0.020(10) \\
$^{46}$V  & 3.57\footnotemark[1] & 4.86 & 0.040 & 0.075 & 0.075(30) \\
$^{50}$Mn & 3.69 & 3.62 & 0.057 & 0.054 & 0.035(20) \\
$^{54}$Co & 2.56 & 2.26 & 0.058 & 0.045 & 0.050(30) \\[3mm]
$^{42}$Sc $-$ $^{42}$Ti & & & -0.031 & -0.096 & -0.080(15) \\
$^{46}$V  $-$ $^{46}$Cr & & &  0.028 &  0.052 &  0.030(20) \\
$^{50}$Mn $-$ $^{50}$Fe & & &  0.036 &  0.035 &  0.010(15) \\
$^{54}$Co $-$ $^{54}$Ni & & &  0.028 &  0.022 & -0.015(60) \\[-3mm]
\footnotetext[1]{Second excited $0^+$state; shell-model calculations indicate
this state takes up most of the depletion from the analog state.}
\end{tabular}
\end{ruledtabular}
\end{table*}

The details of the calculations for $\delta_{C1}$ are described in Ref.~\cite{TH08}.  If isospin were an exact symmetry then
the decay of the parent $0^+, T=1$ state would proceed exclusively to its $0^+$ analog state in the daughter nucleus.  Fermi
transitions to all other $0^+$ states in the daughter would be expressly forbidden.  But when charge-dependent terms are added
to the shell-model Hamiltonian there is some depletion of the analog transition strength, with the missing strength appearing in
weak transitions to excited $0^+$ states.  Significantly, in many cases the bulk of the analog-state depletion shows up in feeding
a single excited $0^+$ state, usually (but not always) the lowest excited one.  In the limit of two-state mixing, perturbation
theory would indicate that
\be
\delta_{C1} \propto \frac{1}{(\Delta E)^2}
\label{dC1DE}
\ee
where $\Delta E$ is the energy separation of the analog and non-analog $0^+$ states.  Since the calculated energy separation in the
shell model, $(\Delta E)_{\rm theo}$, does not exactly match the experimental value, $(\Delta E)_{\rm expt}$, we refine our model
calculation of $\delta_{C1}$ by scaling its value by a factor $(\Delta E)^2_{\rm theo} / (\Delta E)^2_{\rm expt}$.

Our $\delta_{C1}$ results for the decays of $^{42}$Ti, $^{46}$Cr, $^{50}$Fe and $^{54}$Ni are found in Table~\ref{t:tab3}, where
they can be compared with the mirror decays of $^{42}$Sc, $^{46}$V, $^{50}$Mn and $^{54}$Co.  In each case, columns 2 and 3 give the
experimental and calculated excitation energies of the non-analog $0^+$ state that takes the bulk of the Fermi strength depleted from
the analog states.  Columns 4 and 5 give $\delta_{C1}$ without and with scaling by $(\Delta E)^2_{\rm theo} / (\Delta E)^2_{\rm expt}$.

For each nucleus, we performed several shell-model calculations with several different charge-independent effective Hamiltonians --
the same as those described and referenced in Ref.~\cite{TH08}.  Only one of these calculations is recorded in the table but the
adopted value, which appears in the sixth column, represents an average over all calculations, with an uncertainty assigned to span
the range of results obtained.

\begin{table*}
\caption{Calculations of $\delta_{C2}$ with Woods-Saxon radial functions for three methodologies ($\delta_{C2}^{II}$, 
$\delta_{C2}^{III}$, $\delta_{C2}^{IV}$) for one sample shell-model interaction.  The adopted values and uncertainties reflect
the spread in results for several shell-model interactions, different methodologies, and the uncertainty in the radius parameter,
$r_0$.  The last four lines give the differences in isospin-symmetry-breaking corrections for the four mirror transitions.  Note
that the uncertainties of the mirror differences in $\delta_{C2}$ were not determined from the uncertainties on the two contributing
$\delta_{C2}$ values but were independently evaluated to cover the spread in the calculated {\it differences}.
\label{t:tab4}}
\begin{ruledtabular}
\begin{tabular}{rlldddr}
& & & & & & \\[-3mm]
\multicolumn{1}{r}{Parent} &
\multicolumn{2}{c}{Radius parameters (fm)} & & & &
\multicolumn{1}{r}{Adopted } \\
\multicolumn{1}{r}{nucleus} &  
\multicolumn{1}{r}{$\langle r^2 \rangle^{1/2}$} &  
\multicolumn{1}{c}{$r_0$~~~} &  
\multicolumn{1}{r}{$\delta_{C2}^{II}(\%)$} &  
\multicolumn{1}{r}{$\delta_{C2}^{III}(\%)$} &  
\multicolumn{1}{r}{$\delta_{C2}^{IV}(\%)$} &  
\multicolumn{1}{r}{$\delta_{C2}(\%)$} \\[1mm]  
\hline
& & & & & & \\[-2mm]
\multicolumn{1}{l}{$T_z=-1$} &&&&&&\\
$^{42}$Ti & ~~3.616(5) & ~~1.323(2) & 0.901 & 0.869 & 0.800 & 0.855(60) \\
$^{46}$Cr & ~~3.70(10) & ~~1.316(44) & 0.764 & 0.723 & 0.658 & 0.715(85) \\
$^{50}$Fe & ~~3.58(6) & ~~1.206(24) & 0.674 & 0.613 & 0.615 & 0.635(45) \\
$^{54}$Ni & ~~3.68(5) & ~~1.201(21) & 0.784 & 0.684 & 0.710 & 0.725(60) \\[1mm]
\multicolumn{1}{l}{$T_z=0$} &&&&&&\\
$^{42}$Sc & ~~3.570(24) & ~~1.319(11) & 0.704 & 0.681 & 0.632 & 0.670(45) \\
$^{46}$V  & ~~3.60(7) & ~~1.285(31) & 0.587 & 0.542 & 0.506 & 0.545(55) \\  
$^{50}$Mn & ~~3.712(20) & ~~1.273(8) & 0.657 & 0.621 & 0.615 & 0.630(25) \\
$^{54}$Co & ~~3.83(7) & ~~1.275(29) & 0.760 & 0.688 & 0.706 & 0.720(60) \\[3mm]
$^{42}$Sc $-$ $^{42}$Ti & & & -0.197 & -0.188 & -0.168 & $-0.185(20)$ \\
$^{46}$V  $-$ $^{46}$Cr & & & -0.177 & -0.181 & -0.152 & $-0.170(80)$ \\
$^{50}$Mn $-$ $^{50}$Fe & & & -0.017 &  0.008 &  0.000 & $-0.005(40)$ \\
$^{54}$Co $-$ $^{54}$Ni & & & -0.024 &  0.004 & -0.004 & $-0.005(60)$ \\[1mm]
\end{tabular}
\end{ruledtabular}
\end{table*}

Next, we consider $\delta_{C2}$.  For its computation, the radial functions we use in the overlap integral are eigenfunctions
of a Woods-Saxon potential, as justified in our survey article \cite{HT15}.  The methods of calculation have been described in
detail in \cite{TH08,TH02}.  Much benefit is gained from a very strong constraint: The asymptotic forms of all radial functions
must match the measured separation energies $S_p$ and $S_n$, where $S_p$ is the proton separation energy in the decaying nucleus
and $S_n$ is the neutron separation energy in the daughter nucleus.  The Woods-Saxon potential for a nucleus of mass $A$ and
charge $Z+1$ is taken to be the following:
\be
V(r) = - V_0 f(r) - V_s g(r) {\bf l} \cdot \mbox{\boldmath $\sigma$}
+ V_C(r) - V_g g(r) - V_h h(r)
\label{WS}
\ee
where
\bea
f(r) & = & \left \{ 1 + \exp \left ( \frac{r - R}{a} \right ) \right \}^{-1},
\nonumber \\
g(r) & = & \left ( \frac{\hbar}{m_{\pi} c} \right )^2 \frac{1}{a_s r}
\exp \left ( \frac{r - R_s}{a_s} \right )
\nonumber \\
& & ~ \times \left \{ 1 + \exp \left ( \frac{r - R_s}{a_s} \right ) \right \}^{-2} ,
\nonumber \\
h(r) & = & a^2 \left ( \frac{df}{dr} \right )^2 ,
\nonumber \\
V_C(r) & = & Z e^2/r ~~~~~ {\rm for} ~ r \geq R_c ,
\nonumber \\
& = & \frac{Z e^2}{2  R_c} \left ( 3 - \frac{r^2}{R_c^2} \right ) ~~~~
{\rm for} ~ r < R_c ,
\label{WS1}
\eea
with $R = r_0 (A-1)^{1/3}$ and $R_s = r_s (A-1)^{1/3}$.  The first three terms in Eq.~(\ref{WS}) are the central, spin-orbit and
Coulomb terms respectively.  The fourth and fifth are additional surface terms whose role we discuss shortly.

Most of the parameters are fixed at standard values, $V_s = 7$ MeV, $r_s = 1.1$ fm and $a = a_s = 0.65$ fm, and the radius of the
Coulomb potential, $R_c$, is determined from the root-mean-square charge radius, $\langle r^2 \rangle^{1/2}$, of the decaying nucleus.  
Likewise the radius parameter of the central potential, $r_0$, is determined by requiring that the charge density constructed
from the proton eigenfunctions of the potential yields a root-mean-square charge radius $\langle r^2 \rangle^{1/2}$ in agreement with
the known experimental value.  The radius parameters used in our calculations of $\delta_{C2}$ appear in the second and third columns of
Table~\ref{t:tab4}.

Our results for $\delta_{C2}$ itself, calculated with three different methodologies, are given in columns 4-6 of Table~\ref{t:tab4},
with the ultimately adopted values in column 7. The shell model enters these computations because the initial and final $A$-particle
states are expanded in a complete set of $(A-1)$-particle states and single-particle states.  The shell model provides the expansion
coefficients.  For a state in the $(A-1)$ system at an excitation energy $E_x$, the proton and neutron separation energies assigned
to the single particle for this term in the expansion are $S_p + E_x$ and $S_n + E_x$.  For the methodology labeled $II$, the strength
of the central potential $V_0$ was continually readjusted for each term in the parentage expansion to reproduce these separation
energies.  With the radial overlap integral obtained from these eigenfunctions, the isospin-symmetry-breaking correction is labelled
$\delta_{C2}^{II}$.  Alternatively, the adjustment to the Woods-Saxon potential can be accomplished with the surface terms: For
$\delta_{C2}^{III}$ we adjusted $V_g$ and for $\delta_{C2}^{IV}$ it was $V_h$ that was adjusted.  Further details of this approach
are given in Refs.~\cite{TH08,TH02}. 

In Table~\ref{t:tab4}, the $\delta_{C2}$ results for the three different methodologies are given for one sample shell-model
interaction.  The adopted value is an average over the different shell-model calculations and different methodologies with an
uncertainty that covers the spread in the results and the uncertainty associated with the experimental root-mean-square charge radius. 

\begin{table*}
\caption{Calculated $ft^a/ft^b$ ratios for the four mirror doublets.
\label{t:tab5}}
\begin{ruledtabular}
\begin{tabular}{lrrr}
& & & \\[-3mm]
\multicolumn{1}{c}{Decay pairs $a;b$} &
\multicolumn{1}{r}{$\delta_R^b - \delta_R^a (\%)$} &
\multicolumn{1}{r}{$\delta_C^b - \delta_C^a (\%)$} &
\multicolumn{1}{r}{$ft^a / ft^b$} \\[1mm]
\hline
& & & \\[-1mm]
$^{42}$Ti $\rightarrow$ $^{42}$Sc ; $^{42}$Sc $\rightarrow$ $^{42}$Ca&
0.296(30) & $-0.265(25)$ & 1.00561(39) \\
$^{46}$Cr $\rightarrow$ $^{46}$V  ; $^{46}$V  $\rightarrow$ $^{46}$Ti &
0.165(10) & $-0.140(82)$ & 1.00305(83) \\
$^{50}$Fe $\rightarrow$ $^{50}$Mn ; $^{50}$Mn $\rightarrow$ $^{50}$Cr &
0.120(20) & $ 0.005(43)$ & 1.00115(47) \\
$^{54}$Ni $\rightarrow$ $^{54}$Co ; $^{54}$Co $\rightarrow$ $^{54}$Fe &
0.143(30) & $-0.020(85)$ & 1.00163(90) \\[1mm]
\end{tabular}
\end{ruledtabular}
\end{table*}

The question of what is the appropriate root-mean-square charge radius had to be revisited for these calculations following the
recent compilation of experimental results by Angeli and Marinova \cite{AM13}, which were not incorporated into our 2015 survey
\cite{HT15}.  Considering first the $T_z=0$ parent nuclei, we find that for two of them, $^{46}$V and $^{54}$Co, there have been
no updates in charge radii, so the results given in Table~\ref{t:tab4} for these nuclei are identical to those published in 2008
\cite{TH08} and used in 2015.  However, for $^{42}$Sc and $^{50}$Mn, new experimental charge radii have appeared so the $\delta_{C2}$ values
for these nuclei have had to be recomputed.  Their new $\delta_{C2}$ results, shown in Table~\ref{t:tab4}, are slightly higher than
before and have smaller uncertainties compared to those assigned in 2008, reflecting the greater precision of the new charge radii.
The reduction is limited, though, by contributions from uncertainties arising from the spread in results among the different
methodologies and different shell-model interactions, which remains unchanged from before.  Reassuringly, the new results for
$\delta_{C2}$ agree well with those published in 2008 \cite{TH08} within the latter's stated uncertainties.

As to the $T_z=-1$ parents, charge radii are not known for $^{42}$Ti, $^{46}$Cr, $^{50}$Fe and $^{54}$Ni, although they are for
heavier isotopes of each element, typically for those with masses $A+4$, $A+6$ and $A+8$.  In each case, we have done a quadratic
fit to the known charge radii and then extrapolated four mass units back to the isotope of interest.  A generous error is assigned
to charge radii obtained in this manner.  Table~\ref{t:tab4} lists the root-mean-square charge radii we have used for these nuclei.
For $^{42}$Ti, the new results here represent a modest update to those published in 2008 \cite{TH08}, although the new adopted
value of $\delta_{C2}$ is well within the previous uncertainty.

The last four lines of Table~\ref{t:tab4} give the differences in $\delta_{C2}$ values for the four mirror pairs of transitions.
It is interesting to
observe that the differences for the mass-42 and 46 pairs are about 0.2\%, significantly larger than those for masses 50 and 54,
which are nearly zero.  This is something that could be tested in future higher-precision experiments.

\section{The $\F t$ values}
\label{s:Ft}

In Sec.~\ref{s:expt}, world data were evaluated for transitions from four $T_z=-1$ parent nuclei, and the results were entered
into Table~\ref{t:tabb}, where the equivalent (previously evaluated \cite{HT15}) information for their mirror transitions from
$T_z=0$ nuclei also appear. The derived $ft$ values for all eight transitions are also given.  With the theoretical corrections,
$\delta_R^{\prime}$, $\delta_{NS}$, $\delta_{C1}$ and $\delta_{C2}$ that appear in Tables~\ref{t:tab1}-\ref{t:tab4} respectively,
we are now in a position to use Eq.~(\ref{Ft}) to obtain the $\F t$ values for all eight transitions.  Columns 5 and 6 of
Table~\ref{t:tabb} give the theoretical corrections combined as they appear in Eq.~(\ref{Ft}), and column 7 lists the final
$\F t$ values.

It is well known that the $\F t$ values for superallowed transitions provide valuable tests of weak-interaction physics.  In accordance
with Conservation of the Vector Current (CVC), all the $\F t$ values should be the same irrespective of the particular nuclei in which
they are determined.  Once consistency is established among the measured $\F t$ values, the resulting average $\overline{\F t}$ value
can then be used to determine $\Vud$, the up-down element of the Cabibbo-Kobayashi-Maskawa (CKM) matrix.  The $\Vud$ result is a key
ingredient in the most-definitive available test of CKM matrix unitarity, a fundamental principle of the standard model.

The four $\F t$ values in Table~\ref{t:tabb} for transitions from $T_z=0$ parents have already been incorporated in the most recent
evaluation of $\Vud$ \cite{HT15}.  Their $\sim0.06$\% precision is representative of the 14 transitions used in that evaluation.  Clearly
the four $\F t$ values for the $T_z=-1$ cases currently lack the precision to contribute to this picture.  However, that could change in
future as experimental improvements are made, especially in the measurement of $Q_{EC}$.  For now, though, we can declare that the $\F t$
values for the $T_z=-1$ cases are consistent with the current best value for the average \cite{HT15}: $\overline{\F t} = 3072.27\pm 0.62$~s 

\section{Mirror asymmetry}
\label{s:mirror}

The addition of three new proton-rich $T_z = -1$ $\beta$ emitters whose superallowed Fermi branches are the isospin mirrors of already
well-known $T_z = 0$ $\beta$ emitters gives us the opportunity to examine the ratio of $ft$ values for these mirror transitions and to
discuss their asymmetry in terms of isospin-symmetry breaking.  This approach has already been advanced for the mirror Fermi decays of
$^{38}$Ca and $^{38m}$K by Park \etal \ \cite{Pa14,Pa15}.  If we accept the CVC requirement that all the $T=1$ superallowed transitions
must have the same $\F t$ values, then obviously this requirement applies to each mirror pair and, from Eq.~(\ref{Ft}), we can derive
the following expression for the ratio of experimental $ft$ values for such a pair:
\be 
\frac{ft^a}{ft^b} = 1 + (\delta_R^b - \delta_R^a)
- (\delta_C^b - \delta_C^a) ,
\label{ftft}
\ee
where superscript ``$a$" denotes the decay of the $T_z=-1$ parent and ``$b$" denotes the decay of the mirror $T_z=0$ parent.  Here $\delta_R
= \delta_R^{\prime} + \delta_{NS}$ and $\delta_C = \delta_{C1} + \delta_{C2}$ and their mirror differences are already listed in
Tables~\ref{t:tab1}, \ref{t:tab2}, \ref{t:tab3} and \ref{t:tab4}.  The advantage offered by Eq.~(\ref{ftft}) is that the theoretical
uncertainty on a difference like $\delta_C^b - \delta_C^a$ is less than the uncertainties on $\delta_C^b$ and $\delta_C^a$ individually.

In Table~\ref{t:tab5} we list values of $\delta_R^b - \delta_R^a$ and $\delta_C^b - \delta_C^a$ and hence values for $ft^a / ft^b$.
These values differ from unity by amounts ranging from $0.1 \%$ to $0.6 \%$ with radiative-correction and isospin-symmetry-breaking
differences contributing comparably.  With future experimental precision at the $\sim 0.1 \%$ level, it would become possible to
test the corrections for these pairs in the way first demonstrated by Park \etal \ \cite{Pa14} for the mirror superallowed decays of
$^{38}$Ca and $^{38m}$K.  Particularly attractive is the mass-42 mirror pair, for which the $ft$-value ratio is expected to differ
from unity by nearly 0.6\%.

\begin{table*}
\caption{Values of the coefficients $a_0$ and $a_1$ that yield the statistical rate function $f_0$ from Eq.~(\ref{f0fit}), and
coefficients $b_0$, $b_1$, $b_2$ and $b_3$ that yield the correction $\delta_S$ from Eq.~(\ref{ds2}).  Coefficients $a_2$ and
$a_3$ are held fixed at the values: $a_2 = -2/15$ and $a_3 = 1/4$.  
\label{fparam}}
\begin{ruledtabular}
\begin{tabular}{rrrrrrr}
& & & & & & \\[-3mm]
\multicolumn{1}{r}{Parent} & & & & & & \\
\multicolumn{1}{r}{nucleus}
 & \multicolumn{1}{c}{$a_0$}
 & \multicolumn{1}{c}{$a_1$}
 & \multicolumn{1}{c}{$b_0(\%)$}
 & \multicolumn{1}{c}{$b_1(\%)$}
 & \multicolumn{1}{c}{$b_2(\%)$}
 & \multicolumn{1}{c}{$b_3(\%)$}
 \\[1mm]
\hline
& & & & & & \\[-3mm]
   $^{46}$Cr &                                     
   0.0207203  &  $-0.0797342$  &
   0.29193  &   0.17401  &   0.26989  &  $-0.00219$ \\
   $^{50}$Fe &                                     
   0.0200743  &  $-0.0845341$  &
   0.34970  &   0.17589  &   0.27937  &  $-0.00199$ \\
   $^{54}$Ni  &                                    
   0.0191989  &  $-0.0398293$  &
   0.42003  &   0.20090  &   0.31418  &  $-0.00216$ \\[1mm]
\end{tabular}
\end{ruledtabular}
\end{table*}

\section{Parameterization of $f$ for $^{46}$Cr, $^{50}$Fe and $^{54}$Ni}
\label{fParam}

To hone the $\F t$ values for the decays of $^{46}$Cr, $^{50}$Fe and $^{54}$Ni to the precision required to compete effectively with the currently
well-known superallowed transitions,  the $Q_{EC}$ values in particular will have to be improved considerably.  When this happens, the statistical
rate function, $f$, will have to be calculated with a precision to match.  We recently published \cite{TH15} a parameterization of $f$ that
allows a user to easily calculate the $f$ value to high precision ($\pm$0.01\%) for the 20 transitions included in our survey \cite{HT15}.  For
completeness, we give in Table~\ref{fparam} the parameters required to calculate $f$ for the three transitions we have added here.

We follow the parameterization developed in Ref.~\cite{TH15}, in which
\be
f = f_0 (1 + \delta_S) ,
\label{ffactor}
\ee
where
\be
f_0 = a_0 W_0^4 p_0 + a_1 W_0^2 p_0 + a_2 p_0 + a_3 W_0 \ln (W_0 + p_0) 
\label{f0fit}
\ee
and
\be
\delta_S = b_0 + b_1 W_0 + b_2/W_0 + b_3 W_0^2, 
\label{ds2}
\ee
where $W_0$ is the maximum total energy of the decay positron 
in electron rest-mass units and $p_0 = (W_0^2-1)^{1/2}$ is the corresponding momentum. 
Two of these parameters are fixed: $a_2 = -2/15$ and $a_3 = 1/4$. The other six are listed in Table~\ref{fparam}. 

Note that, as in Ref.~\cite{TH15}, this parameterization is only valid for the transitions identified and only for a limited range of energies
($\pm$60 keV) around the currently accepted $Q_{EC}$ values.

\section{Conclusions}
\label{s:conc}

Prompted by new measurements from Molina \etal \ \cite{Mo15}, we have thoroughly examined the superallowed Fermi decays of $^{42}$Ti,
$^{46}$Cr, $^{50}$Fe and $^{54}$Ni, the latter three of which having never before been included in our periodic surveys of world data for
such decays.

We began this report by assembling all pertinent references and arriving at recommended results for the $Q_{EC}$ values,
half-lives and branching ratios for all four transitions; next, we presented calculations of their radiative and isospin-symmetry-breaking
corrections.  From this input we obtained their $ft$ and $\F t$ values.

The results have all been presented in such a way that these four
transitions from $T_z=-1$ nuclei could be compared with their mirror superallowed transitions from the $T_z=0$ nuclei $^{42}$Sc, $^{46}$V,
$^{50}$Mn and $^{54}$Co.  This also gave us the opportunity to update the $\delta_C$ values for $^{42}$Ti, $^{42}$Sc and $^{50}$Mn in order
to incorporate an update in the recommended values for the root-mean-square charge radii, $\langle r^2 \rangle^{1/2}$, of these nuclei as
tabulated by Angeli and Marinova \cite{AM13}.

By presenting our results in terms of comparisons of mirror pairs of transitions with $A$=42, 46, 50 and 54, we demonstrate
the importance of measuring the $T_z=-1$ members of these mirror pairs with improved precision.  The difference in the $ft$ values between
the two members of each mirror pair is sensitive to the calculated correction terms, and can be used to test, and possibly improve, them. 

Although the $ft$-value uncertainties for the decays of $^{42}$Ti, $^{46}$Cr, $^{50}$Fe and $^{54}$Ni are still too large for this purpose,
we take the view that, with experimental accessibility now demonstrated, there is sufficient motivation to proceed with improving the precision.
An obvious place to begin is with modern re-measurements of the 40-year-old $Q_{EC}$ values for the decays of $^{46}$Cr, $^{50}$Fe and
$^{54}$Ni with a precision to match the recent Penning-trap measurement of the $^{42}$Ti $Q_{EC}$ value.

To aid in this endeavor, we have also provided the means to easily calculate $f$ values for the superallowed transitions from $^{46}$Cr, $^{50}$Fe
and $^{54}$Ni to the required $\pm$0.01\% precision.

\begin{acknowledgments}

This material is based upon work supported by the U.S. Department of Energy, Office of Science, Office of Nuclear Physics, under
Award Number DE-FG03-93ER40773, and by the Robert A. Welch Foundation under Grant No.\,A-1397.

\end{acknowledgments}


\begin{thebibliography}{99999}


\bibitem{HT15}
J.C. Hardy and I.S. Towner, \pr C {\bf 91}, 025501 (2015).

\bibitem{Mo15}
F. Molina \etal , \pr C {\bf 91}, 014301 (2015).

\bibitem{TH10}
I.S. Towner and J.C. Hardy, \pr C {\bf 82}, 065501 (2010).

\bibitem{TH15}
I.S. Towner and J.C. Hardy, \pr C {\bf 91}, 015501 (2015).

\bibitem{En90}
P.M. Endt, Nucl. Phys. {\bf A521}, 1 (1990).

\bibitem{Ga69}
A Gallmann, E. Aslanides, F. Jundt and E. Jacobs, Phys. Rev. {\bf 186}, 1160 (1969).

\bibitem{Ku09}
T. Kurtukian Nieto, J. Souin, T. Eronen, L. Audirac, J. Aysto, B. Blank, V.-V. Elomaa, J. Giovinazzo, U. Hager, J. Hakala, A. Jokinen, A. Kankainen, P. Karvonen, T. Kessler, I.D. Moore, H. Penttila, S. Rahaman, M. Reponen, S. Rinta-Antila, J. Rissanen, A. Saastamoinen, T. Sonoda and C. Weber, Phys. Rev. C {\bf 80}, 035502 (2009).

\bibitem{On05}
T.K. Onishi, A. Gelberg, H. Sakurai, K. Yoneda, N. Aoi, N. Imai, H. Baba, P. von Brentano, N. Fukuda, Y. Ishikawa, M. Ishihara, H. Iwasaki, D. Kameda, T. Kishida, A.F. Lisetskiy, H.J. Ong, M. Osada, T. Otsuka, M.K. Suzuki, K. Ue, Y. Utsuno and H. Watanabe, Phys. Rev. C {\bf 72}, 024308 (2005).

\bibitem{Re99}
I. Reusen, A. Andreyev, J. Andrzejewski, N. Bijnens, S. Franchoo, M. Huyse, Yu. Kudryavtsev, K. Kruglov, W.F. Mueller, A. Piechaczek, R. Raabe, K. Rykaczewski, J. Szerypo, P. Van Duppen, L. Vermeeren, J. Wauters and A. Wöhr, Phys. Rev. C {\bf 72}, 024308 (2005).

\bibitem{Tr77}
R.E. Tribble, J.D. Cossairt, D.P. May and R.A. Kenefick, Phys. Rev. C {\bf 16}, 917 (1977).

\bibitem{Wa12}
M. Wang, G. Audi, A.H. Wapstra, F.G. Kondev, M. MacCormick, X. Xu and B. Pfeiffer, Chinese Physics C {\bf 36}, 1603 (2012)

\bibitem{Zi72}
J. Zioni, A.A. Jaffe, E. Friedman, N. Haik, R. Schectman and D. Nir, Nucl. Phys. {\bf A181}, 465 (1972).

\bibitem{TH08}
I.S. Towner and J.C. Hardy, \pr C {\bf 77}, 025501 (2008).

\bibitem{Si67}
A. Sirlin, \pr {\bf 164}, 1767 (1967).

\bibitem{TH02}
I.S. Towner and J.C. Hardy, \pr C {\bf 66}, 035501 (2002).

\bibitem{To94}
I.S. Towner, \pl {\bf B333}, 13 (1994).

\bibitem{AM13}
I. Angeli and K.P. Marinova, Atomic Data and Nuclear Data Tables
{\bf 99}, 69 (2013).

\bibitem{Pa14}
H.I. Park, J.C. Hardy, V.E. Iacob, M. Bencomo, L. Chen, V. Horvat, N.Nica,
B.T. Roeder, E. Simmons, R.E. Tribble and I.S.Towner,
\prl {\bf 112}, 102502 (2014).

\bibitem{Pa15}
H.I. Park, J.C. Hardy, V.E. Iacob, M. Bencomo, L. Chen, V. Horvat, N.Nica,
B.T. Roeder, E. McCleskey, R.E. Tribble and I.S.Towner,
\pr C {\bf 92}, 015502 (2015).





\end{thebibliography}
\end{document}